# Premature Unilateral Ripening in Euonymus alatus – Two Hits Leave(s) a Red Face

Rithika Narayan[*,1] and Prakash Narayan[1], PhD

* = corresponding author (9th Grader at Elwood JGHS, NY)

1 = 17 Elgar St, East Northport NY-11731

Email: narayanrithika@gmail.com

**Article Type**: *Rapid Communication*

**Key Words**: Euonymus alatus, fall colors, shedding, soil temperature, scale insect

**Dedication** – This work is dedicated to "Papa Joe" Mir, a staunch supporter of science and scientific personnel.

**Acknowledgements** – Dr. Dong-sung Lim of Angion Biomedica NY and Mr. Tom Schmeelk (NYS)

*Rapid Communication*

*Background*

The row of Euonymus alatus a.k.a. burning bushes located at 40⁰52'26.25"N and 73⁰19'28.42"W typically changes color from green to a vivid red starting in the 1st or 2nd week of October. The long axis of this row aligns with the East-West direction (Figure 1) such that the two faces or major/bulk aspects of these bushes are oriented North (PJ) and South (RN).

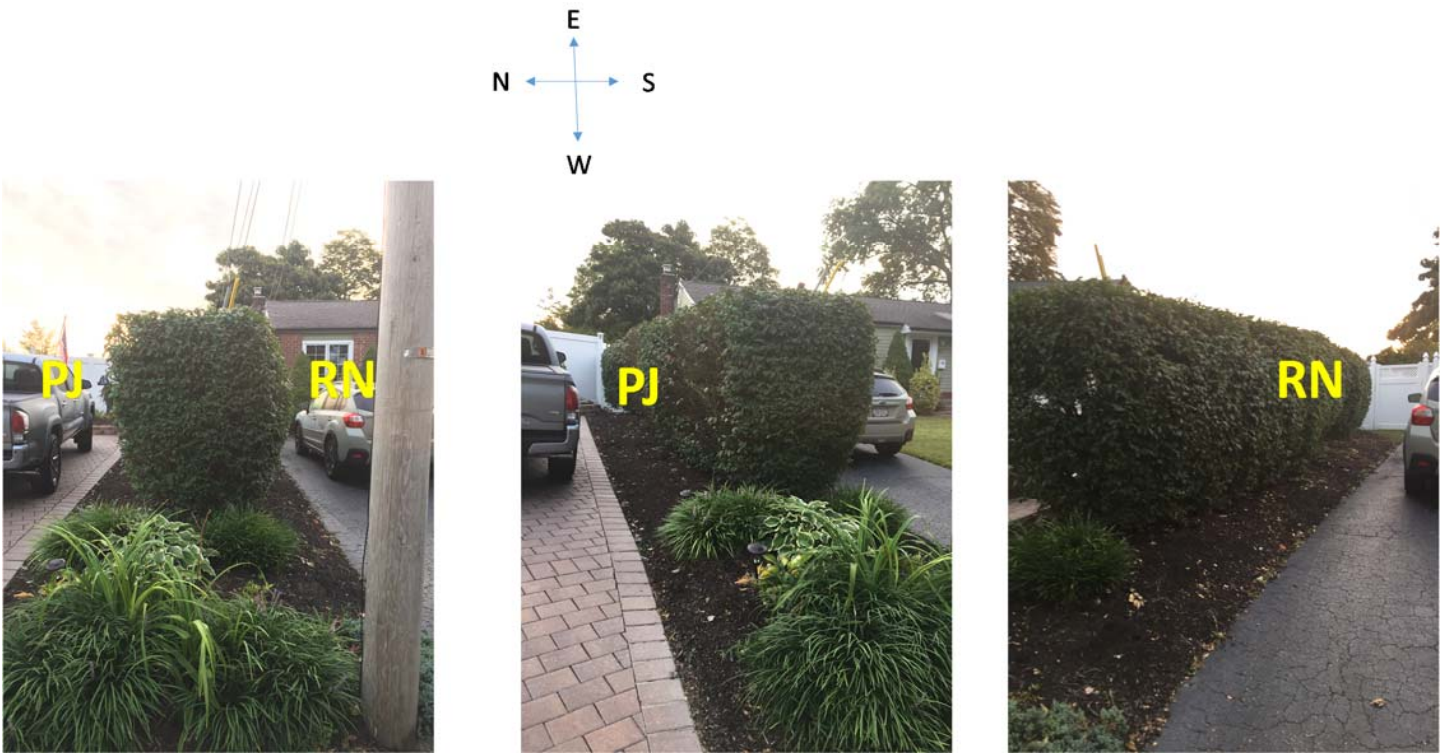

Figure 1. A Row of Euonymus alatus with the long axis aligned in the East-West direction. The Northern aspect is termed PJ and the Southern aspect RN.

This year, i.e. 2017, it was observed that the PJ (North-facing) aspect of these bushes had started turning red by mid-August, ~ 2 months ahead of schedule (Figure 2). By contrast, the RN (South-facing) aspect of the bushes retained their green foliage consistent with historical (across 12 years) observations. By late August, the PJ aspect of the bushes had started shedding its leaves. Noticing that the summer temperatures this year were unseasonably mild, it was hypothesized that an acute temperature gradient across the RN vs. PJ aspects of these bushes was responsible, at least, in part, for the unilateral ripening observed. Specifically, it was posited that a relatively mild summer coupled with the lack of exposure to sunshine resulted in a significantly cooler temperature at the PJ aspect sufficient to trigger leaf ripening.

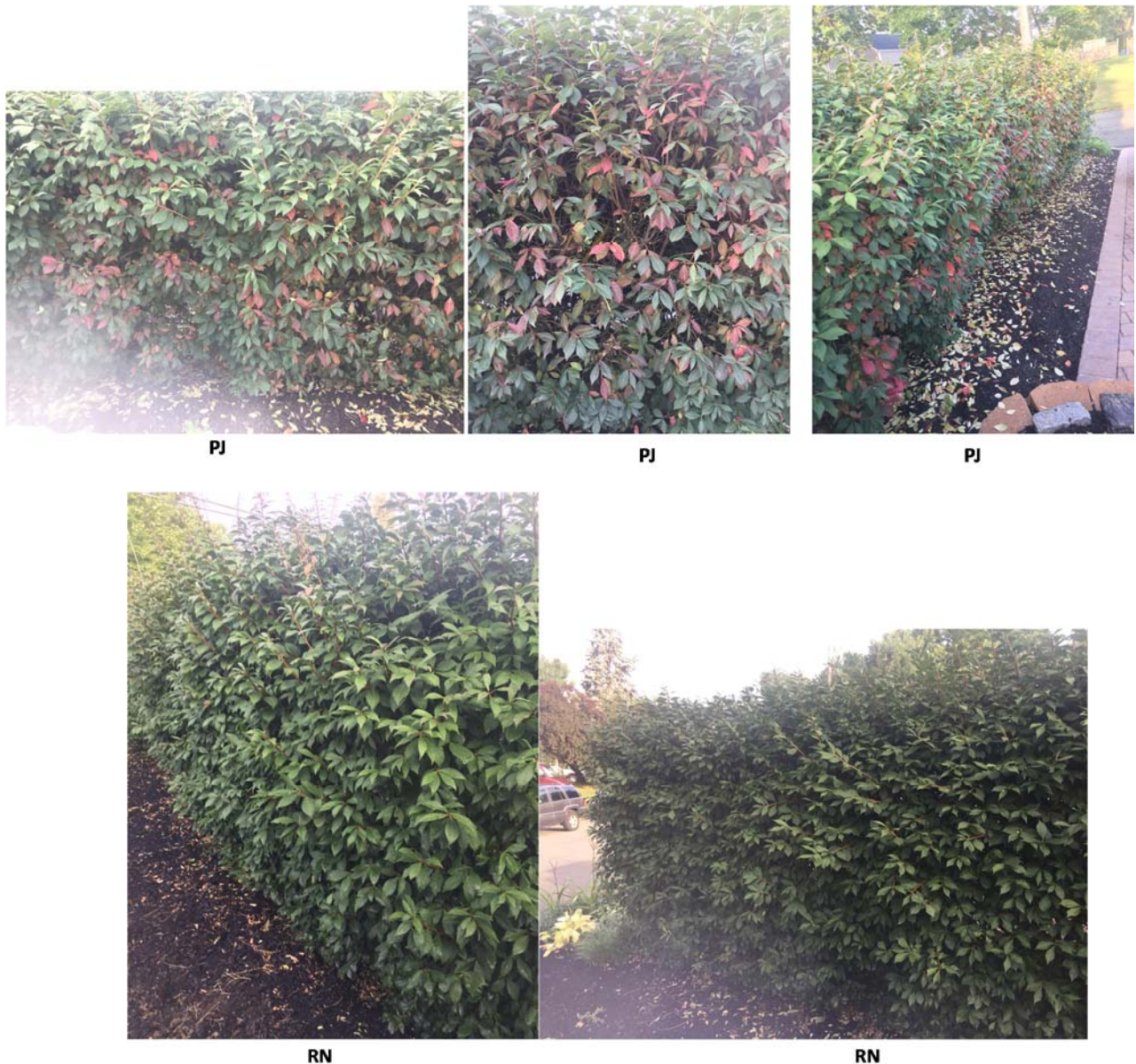

Figure 2. A fall-like pattern was observed by mid to late August in the PJ aspect of the bushes whereas the RN aspect was unremarkable.

*Materials and Methods*

Two Acurite Digital humidity and temperature readers

One WR® Calibrated Electronic Stainless Steel Stem Thermometer

iPhone camera

For nine days that fell within a two week period, temperatures were collected in the morning and evening. The two readers (Figure 3) were calibrated before each reading and temperature (EQ T F) and relative humidity (EQ RH %) were noted. Then, one reader was placed on the soil of the PJ aspect of the bushes while the other was placed on the soil at the opposite RN aspect. Five minutes later, the soil surface temperature (Ambient T F) and relative humidity (RH %) were recorded with RH % serving as a surrogate for soil moisture content. The data were recorded between 7:30 and 9:50 a.m. in the mornings and 3:30 and 7:00 p.m. in the evenings. In addition to collecting soil surface temperatures, internal soil temperatures at a depth of 10-13 cm were obtained for four days at both aspects of the bushes. Whenever possible, internal soil temperatures (soil T C and soil T F) were also obtained twice daily between the hours stated above. In order to determine leaf density, three photographs were obtained from an equivalent distance from either aspect of the bush. A grid was overlaid onto these pictures, and the number of leaves in the center square of the grid was counted. Differences between groups were analyzed using a T-test with a $p < 0.05$ being considered significant.

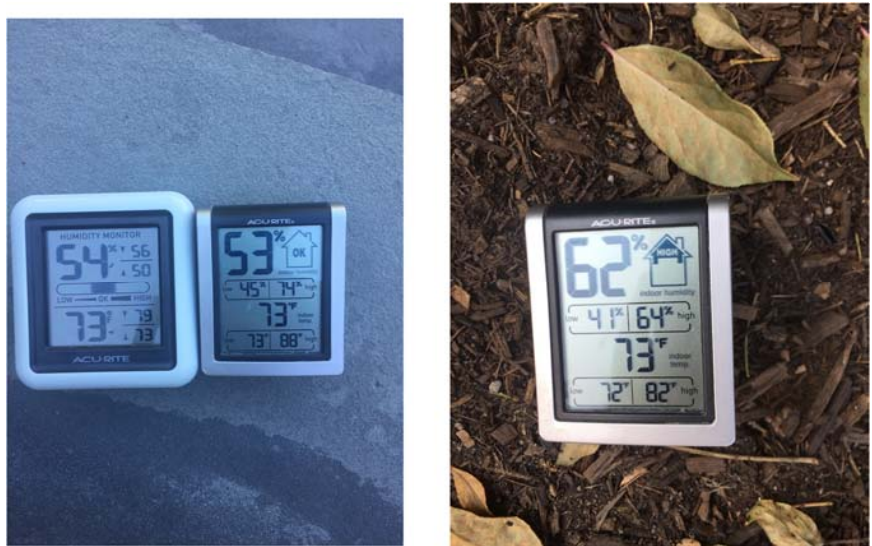

Figure 3. Temperature and relative humidity recordings. The photo on the left shows the recordings during the equilibration protocol. The photo on the right shows soil surface recordings.

*Results*

As seen in Table 1, while there was no difference in the equilibrium readings for temperature (EQTF) or RH (EQ RH %) suggesting that the instruments were functioning satisfactorily. Nevertheless, soil surface temperature (Ambient T F) between the two aspects of the bushes was significantly different ( p = 0.004) during the recorded period. While the RN aspect averaged 89 degrees F, the PJ aspect averaged only 79 degrees F during this period. There was no difference in soil moisture content, in as much as RH % is a surrogate for the same, between the 2 aspects of the bushes.

| Date | Time (Hrs) | EQ T F PJ | EQ T F RN | Ambient T F PJ | Ambient T F RN | | EQ RH % PJ | EQ RH % RN | RH % PJ | RH % RN |
|---|---|---|---|---|---|---|---|---|---|---|
| 8/17/2017 | 1857 | 79 | 79 | 79 | 79 | | 50 | 53 | 55 | 55 |
| 8/18/2017 | 731 | 75 | 75 | 75 | 77 | | 66 | 64 | 74 | 70 |
| 8/18/2017 | 1558 | 79 | 79 | 81 | 79 | | 61 | 64 | 75 | 77 |
| 8/19/2017 | 953 | 75 | 75 | 79 | 97 | | 66 | 65 | 71 | 61 |
| 8/19/2017 | 1548 | 81 | 81 | 88 | 99 | | 59 | 59 | 54 | 55 |
| 8/20/2017 | 908 | 73 | 73 | 73 | 86 | | 53 | 54 | 62 | 55 |
| 8/20/2017 | 1601 | 79 | 79 | 82 | 100 | | 53 | 56 | 44 | 35 |
| 8/23/2017 | 1640 | 82 | 82 | 81 | 84 | | 40 | 52 | 41 | 46 |
| 8/24/2017 | 1536 | 106 | 102 | 90 | 111 | | 16 | 30 | 25 | 24 |
| 8/25/2017 | 1721 | 77 | 77 | 77 | 79 | | 45 | 53 | 44 | 50 |
| 8/26/2017 | 1547 | 77 | 77 | 79 | 99 | | 42 | 50 | 34 | 35 |
| 8/27/2017 | 926 | 70 | 70 | 70 | 81 | | 55 | 59 | 60 | 59 |
| 8/27/2017 | 1538 | 77 | 77 | 79 | 91 | | 46 | 52 | 37 | 40 |
| | Average | 79.23 | 78.92 | 79.46 | 89.38 | | 50.15 | 54.69 | 52.00 | 50.92 |
| | SEM | 2.40 | 2.12 | 1.50 | 3.01 | | 3.68 | 2.49 | 4.44 | 4.11 |
| | P value | | 0.92 | | 0.004 | | | 0.32 | | 0.86 |

Table 1. Soil surface temperature and relative humidity recordings on the PJ and RN aspects of the bushes.

Temperatures recorded on a given day were averaged a plot of temperature vs. day was constructed. The RN aspect had a 13% greater area under curve (AUC) of the temperature.day plot vs. the PJ aspect (Figure 4).

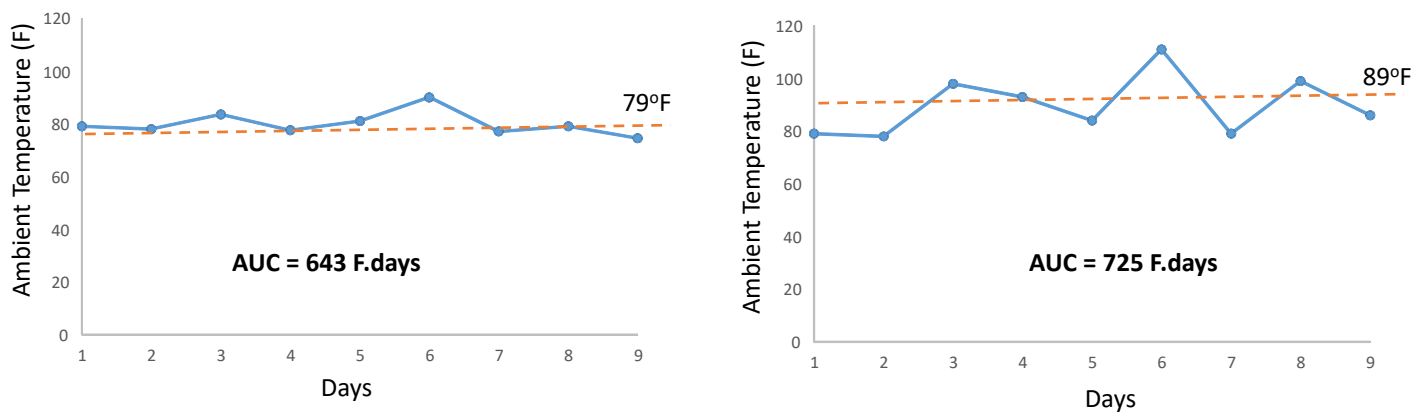

Figure 4. Soil surface AUC for temperature vs. day plots for PJ (left) and RN (right) aspects of the bushes.

Analysis (Figure 5) of the internal soil temperature (Soil T F) also showed a similar pattern with the PJ aspect significantly cooler (72 F; 216 F.days) than the RN aspect (79 F; 236 F.days).

| Date | Soil T C PJ | Soil T C RN | Soil T F PJ | Soil T F RN |
|---|---|---|---|---|
| 8/31/2017 | 23.1 | 25.1 | 74 | 77 |
| 8/31/2017 | 23.2 | 28.4 | 74 | 83 |
| 9/1/2017 | 22.9 | 25 | 73 | 77 |
| 9/1/2017 | 21.8 | 28.8 | 71 | 84 |
| 9/2/2017 | 21.1 | 23.8 | 70 | 75 |
| 9/2/2017 | 21.7 | 24.3 | 71 | 76 |
| 9/3/2017 | 22.7 | 26.1 | 73 | 79 |
| | | Average | 72.24 | 78.67 |
| | | SEM | 0.56 | 1.34 |
| | | P value | | 0.0004 |

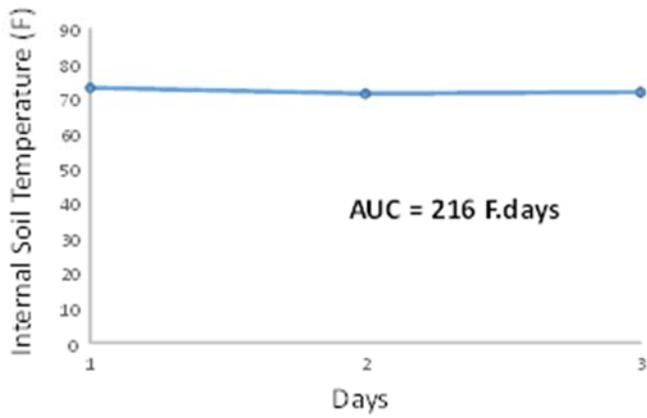
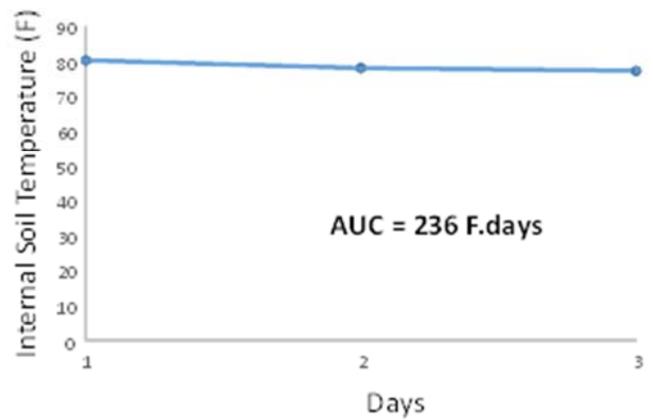

Figure 5. Internal soil temperatures and AUC for temperature vs. day plots for PJ (left) and RN (right) aspects of the bushes.

A recording (Figure 6) of solar movement during this period showed that between 0615 and 1920 hours, the sun spent time only on the RN aspect of the bushes.

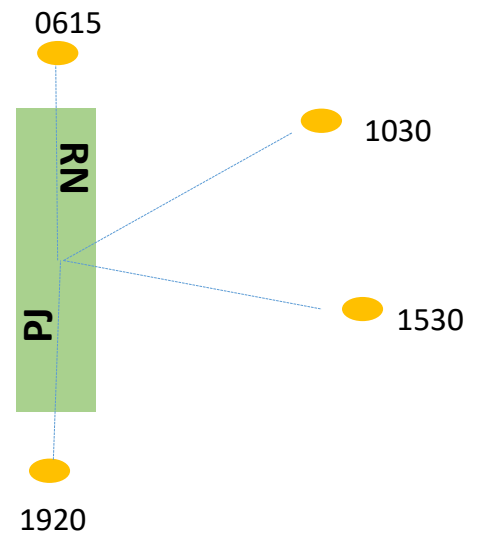

Figure 6. Solar trajectory in late August with respect to the row of bushes.

To determine whether an additional factor such as disease might play a role in these findings, samples from both the PJ and RN aspects were sent to the Department of Environmental Conservation, Delmar, NY. Analysis of the samples revealed the presence (Figure 7) of scale insect (order Hemiptera; suborder Sternorrhyncha; superfamily Coccoidea) on both aspects of the bush with the PJ aspect being relatively more infested.

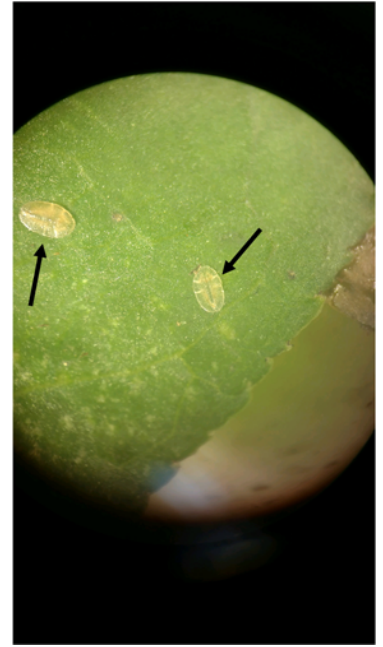

Figure 7. Scale insects (arrows) on the bushes.

By the first week of October, 2017, few, if any, leaves were present on PJ aspect whereas leaves on the RN aspect were only then starting to turn red and thinning out. A very significant difference was observed in the leaf density between these two aspects at this time (Figure 8).

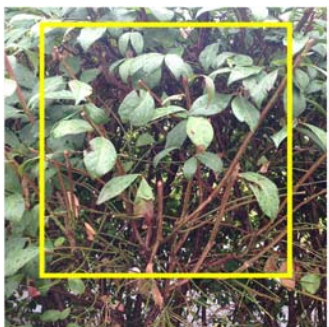
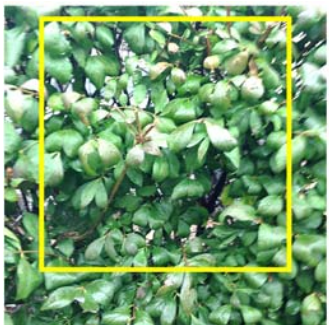
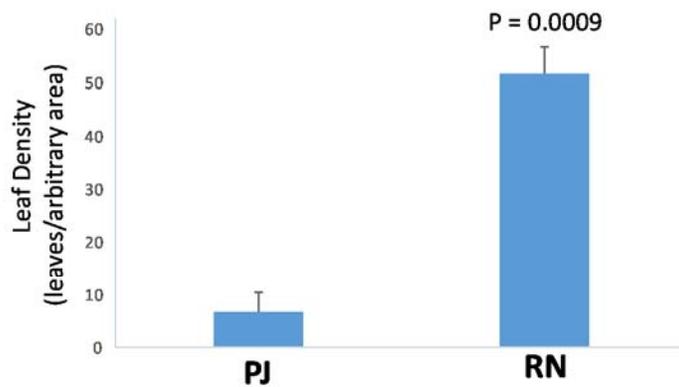

Figure 8. Leaf density at the PJ and RN aspects of the bushes.

*Discussion*

Premature unilateral ripening of a row of Euonymus alatus bushes was observed. Ripening was characterized by reddening and dropping of leaves characteristic of mid to late fall.  The aspect of the bushes that experienced these changes appeared to be exposed to cooler temperatures and higher scale insect infestation.

Fall colors are observed annually in certain trees and bushes in North America typically in early to mid-October followed by shedding of foliage. Shorter periods of sunlight and cooler temperatures result in first a decrease in followed by a cessation of cholorophyll production. A reduction in the green color of chlorophyll  brings out the yellow, orange, red and brown colors of endogenous carotenoids and anthocyanins.  Leaves are finally walled off and the tree/shrub goes into a period of winter hibernation (1,2). A relatively mild summer coupled with cooler- soil surface and -internal soil temperatures and a lack of direct sunlight observed at the PJ aspect of these bushes appear to have conspired to result in the reddening of the leaves on that side. Scale insect infestation might have contributed to a more rapid pathology on that aspect of the bushes. Scale insects are parasitic and subsist on the sap within trees (3). Although these insects were present on both aspects of the bushes, their density was higher on the PJ aspect.

These data suggest that a combination of two etiologically distinct "hits" resulted in the observed phenomenon of unilateral ripening in Euonymus alatus.